\documentclass[11pt,twoside]{article}
\usepackage{graphics,graphicx}
\usepackage{./asp2014}

\newcommand{\comment}[1]{}  
\def\be{\begin{enumerate}}
\def\ee{\end{enumerate}}
\def\bi{\begin{itemize}}
\def\ei{\end{itemize}}
\def\gtrsim{\mathrel{\hbox{\rlap{\hbox{\lower4pt\hbox{$\sim$}}}\hbox{$>$}}}}
\def\lesssim{\mathrel{\hbox{\rlap{\hbox{\lower4pt\hbox{$\sim$}}}\hbox{$<$}}}}
\aspSuppressVolSlug
\resetcounters

\begin{document}
\title{Science with an ngVLA: High-resolution Imaging of Radio Jets Launched by AGN}

\author{Matthew L. Lister,$^1$ Kenneth I. Kellermann,$^2$ and Preeti Kharb$^3$
\affil{$^1$Dept. of Physics and Astronomy, Purdue University, West Lafayette, IN, USA; \email{mlister@purdue.edu}}
\affil{$^2$NRAO, Charlottesville, VA, USA; \email{kkellerm@nrao.edu}} 
\affil{$^3$NCRA-TIFR, Pune, India; \email{kharb@ncra.tifr.res.in}}}


\paperauthor{Matthew L. Lister}{mlister@purdue.edu}{}{Purdue University}{Department of Physics and Astronomy}{West Lafayette}{IN}{
47906}{USA}

\paperauthor{Kenneth I. Kellermann}{kkellerm@nrao.edu}{}{NRAO}{}{Charlottesville}{VA}{}{USA}
\paperauthor{Preeti Kharb}{kharb@ncra.tifr.res.in}{}{NCRA-TIFR}{}{Pune}{}{}{India}

\section{Bridging AGN Jet Studies from Parsec through Kiloparsec Scales}
Jetted plasma outflows from active galactic nuclei (AGN) represent the
most energetic phenomena in the known universe, and play a
key role in regulating galaxy formation through feedback processes \citep{2018NatAs...2..198H}. The JVLA and VLBA have played an
indispensable role in understanding the physics of these powerful jets
and their environments, via high angular resolution full polarization
imaging and astrometric studies. By bridging the current
interferometric gap between the JVLA and VLBA with intermediate
baselines, the ngVLA offers exciting new opportunities to explore the
intermediate regions downstream of the high Lorentz factor pc-scale
AGN jets imaged by VLBI, where entrainment, deceleration, collimation
and particle acceleration all take place.  The ability to image
exceedingly faint radio emission on scales of 10s to 100s of
milliarcseconds will lead to new breakthroughs in resolving some of
the most pressing questions regarding the formation, structure, and
evolution of AGN jets.  These are in turn crucial for a more
complete understanding of the formation of supermassive black holes
and galaxies in the early universe and their subsequent evolution.

\section{Magnetic field structure}
\begin{figure}[h]
  \begin{center}$
\begin{array}{cc}
    \includegraphics[width = 3.5in,angle = 0]{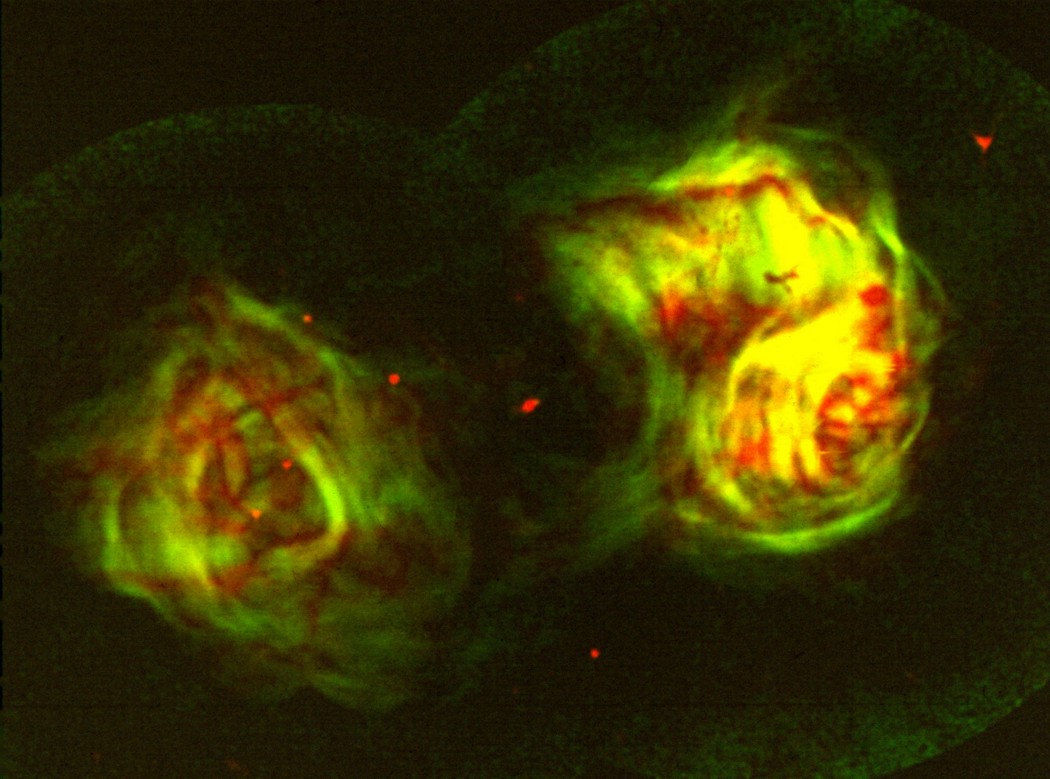} &

\end{array}$
\end{center}

\caption{\label{FornaxA}VLA image of the radio galaxy Fornax A.  In this
  hue-intensity image the brightness corresponds to the intensity of
  the lobes while the highly polarized structure is saturated white
  and regions of increasing depolarization are shaded red. Image
  courtesy of NRAO/AUI.  
}
\end{figure}

Since AGN jets are fundamentally magnetically-driven phenomena,
polarimetric observations provide numerous physical insights otherwise
unattainable via total intensity imaging. As we describe in the
specific science topic sections below, the full (linear and circular)
polarization capabilities of the ngVLA represent a major advance over
existing instruments, as applied to AGN jet studies. Specifically,
Faraday depth, rotation, and depolarization measurements offer a
direct means of probing the properties of jet plasma and its
surrounding medium, as well as the overall magnetic field. VLBA
polarimetric imaging has revealed complex pc-scale magnetic phenomena,
such as MHD waves \citep{2014ApJ...787..151C}, stationary features
\citep{2017ApJ...846...98J}, and structures consistent with planar and
conical shocks \citep{2008Natur.452..966M}. On kiloparsec scales,
polarization studies of radio lobes (Fig.~\ref{FornaxA}) have been a
crucial element supporting the unification of radio galaxies and
quasars \citep{1988Natur.331..147G}, and provide one of the only means
of studying large scale magnetic fields in galaxy clusters
\citep{2010AA...513A..30B}. Although difficult to measure with
existing radio telescope arrays, the detections of circular
polarization \citep{2006AJ....131.1262H} and Faraday rotation
gradients \citep{2017MNRAS.472.1792G} in a few pc-scale
AGN jets have revised our views of internal jet physics over the last
decade, and have prompted many detailed numerical studies to better
understand the jet launching and collimation process
\citep[e.g.,][]{2010ApJ...725..750B,2016MNRAS.461L..46T}. With the new
50-250 km baselines and substantially improved sensitivity afforded by
the ngVLA, it will be possible to study the magnetic field structure
of lobes and terminal jet hotspots in AGN in unprecedented detail.
This will lead to a better understanding of particle acceleration
mechanisms and how relativistic jet flows are decelerated.

\comment{
\bigskip Possible figures:
\be
\item VLA Faraday rotation/gradient map of large radio lobe source:
  fig 2 of

http://adsabs.harvard.edu/abs/2002ARA

\item VLBA image: Faraday gradient in a transversely resolved jet
Figure 5 of

 http://iopscience.iop.org/article/10.1088/0004-637X/733/1/11



\ee
}
\section{Launching and Collimation Mechanisms}

Current theory and numerical simulations for the launching of jets
favor scenarios that typically involve a supermassive black hole and
accretion disk in the nucleus of the host galaxy.  Magnetic fields
play an important role, due to the large kinetic power of the
outflows, and the lack of sufficient thermal and radiation pressure in
the accretion disk region. A toroidal field component can provide a
means of producing the highly collimated jets
imaged by VLBI. Magnetic fields also provide the means of extracting
energy and angular momentum from the accretion disk and a rotating
black hole.

High resolution radio imaging of nearby AGN jets has revealed
important details regarding collimation near the base of the jet. In
the case of the nearby AGN M87, the jet initially has a wide opening
angle, which is sharply collimated at $\sim 100$ Schwarzschild radii
($R_s$) to a roughly parabolic profile \citep{2013ApJ...775...70H}.
Only at $\sim 10^5$ $R_s$ does the jet take on a conical shape
expected for simple adiabatic expansion \citep{2012ApJ...745L..28A}.
Some magnetic launching models \citep[e.g.,][]{2007MNRAS.380...51K}
predict strong collimation once the jet flow passes through the
fast-magnetosonic point, when the bulk conversion of magnetic to
kinetic energy takes place. Multi-epoch VLBI imaging of the inner M87
jet has also shown an evolution of the flow velocity that can be
explained in the framework of MHD jet acceleration and this type of
energy conversion \citep{2016AA...595A..54M,2018ApJ...855..128W}.
Since M87 is unusually close (16 Mpc), the ngVLA will be the only
instrument capable of probing faint structure in the jet further
downstream, including the intriguing HST-1 knot region, which has been
seen to exhibit much higher internal proper motions (up to $\sim 4$ c)
than the (sub-luminal) inner jet \citep{2012AA...538L..10G}. This type
of analysis can also be extended to many other nearby radio galaxies
such as NGC 6251 \citep{2016ApJ...833..288T} and PKS 0227-369
\citep{2013MNRAS.429.1189P}. At present, the only instrument capable
of probing these scales is e-MERLIN, which, having only 7 antennas,
cannot provide the dual-sensitivity needed to probe the diffuse and
compact structures that are present in these nearby jets. The ngVLA
will provide 10 times the continuum imaging sensitivity of e-MERLIN at
5 GHz in one tenth the integration time.

\comment{
\bigskip Possible figures:
\be
\item Mertens M87 velocity field: Fig 2 of

  http://adsabs.harvard.edu/abs/2016A\&A...595A..54M

\ee
}

\section{Jet Structure}
\begin{figure}[h]
  \begin{center}$
    \begin{array}{c}
      \includegraphics[width = 5in,angle = 0]{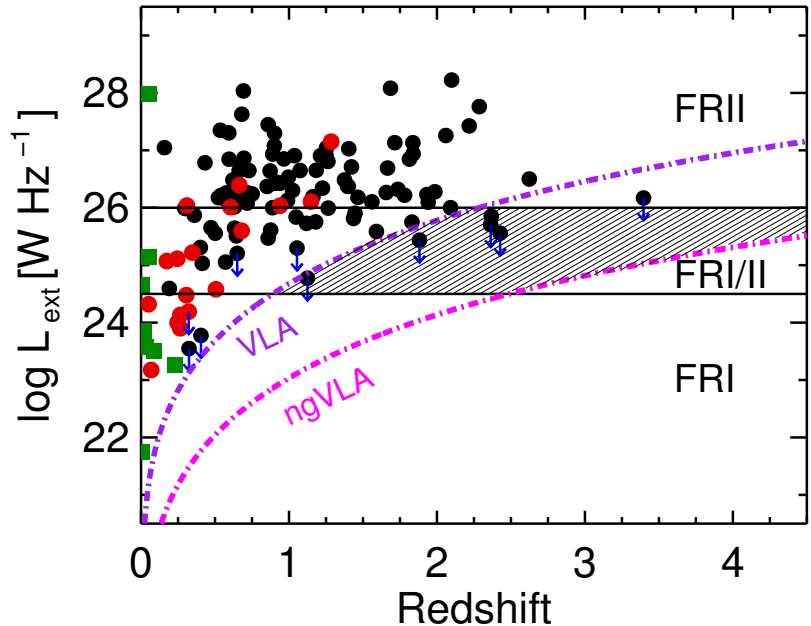} 
      
    \end{array}$
    \end{center}

\caption{\label{Kharbfig1}The 1.4~GHz extended luminosity with respect
  to redshift for the original MOJAVE sample \citep{2010ApJ...710..764K}. Black
  and red circles denote quasars and BL Lac objects, respectively,
  while green squares denote radio galaxies. Core-only sources are
  represented as upper limits with downward arrows. The solid lines
  indicate the FRI$-$FRII divide (extrapolated from 178 MHz to 1.4 GHz
  assuming a spectral index, $\alpha$=0.8), following
  \citet{1996AJ....112....9L} and \citet{2006ApJ...637..183L}. The purple line denotes
  the sensitivity limit for the historical VLA which was used to carry
  out the MOJAVE study, while the magenta line denotes the sensitivity
  limit for the ngVLA (see http://ngvla.nrao.edu/page/refdesign).
  ngVLA will be able to detect more than twice as many ``hybrid
  FRI/II'' sources, compared to previous studies, revolutionising the
  study of radio-loud AGN.}
\end{figure}

\begin{figure}[h]
  \begin{center}$
\begin{array}{cc}
    \includegraphics[width = 3.5in,angle = 0]{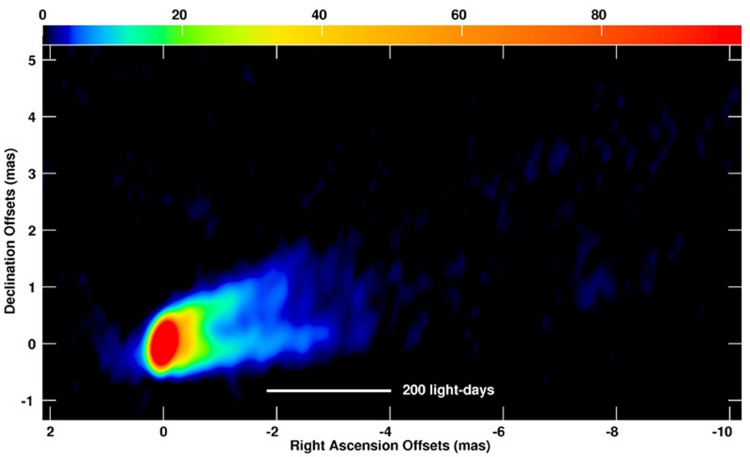} &

\end{array}$
\end{center}

\caption{\label{M87jet}VLBA image of the radio jet in M87. With a high resolution of
  430 by 210 microarcseconds (1.3 by 0.64 light-months), it reveals that the
  jet appears brighter at its edge compared to the center, and
  collimation of the jet is occurring within 100 Schwarzschild radii. Image
  courtesy of NRAO/AUI.  
}
\end{figure}

With its unprecedented sensitivity and baselines out to 1000 km, the
ngVLA holds the exciting prospect of transversely resolving for the
first time the jets of many nearby radio galaxies such as NGC 315, 3C
66B, 3C 264, and NGC 4261 out to large projected distances. Since
radio galaxy jets are oriented closer to the plane of the sky, they
suffer less from projection and relativistic beaming effects. In a
handful of cases with current instruments, the (de-boosted) receding
jet can be imaged and compared with the approaching jet, giving an
accurate measurement of the viewing angle, which is often impossible
to determine by other means.  This technique also provides detailed
information on the velocity field and magnetic field configuration in
the jets, and has been used extensively by \cite{2014MNRAS.437.3405L}
to make important discoveries regarding the accelerations of jet
plasma and its interaction with the external medium at the jet
boundary. Although currently limited to slower and less beamed (FR-I
class) jets, with the ngVLA it will be possible to study the receding
jets and velocity fields in the most powerful (FR-II class) AGN (Figure~\ref{Kharbfig1}). In
particular, more examples of enhanced radio emission at the jet
boundaries will be found, which have been studied so far in only the
nearest AGN jets such as M87 (Fig.~\ref{M87jet}), Mrk 501
\citep{2008AA...488..905G}, and 3C 84 \citep{2014ApJ...785...53N}. At
present it is unclear whether this phenomenon is a result of
interaction with the external medium, or differential beaming
associated with a transverse jet velocity profile, a corkscrew-like
rotation of the flow \citep{2018ApJ...855..128W} or helical magnetic
fields \citep{2011MNRAS.415.2081C}.

\comment{
\bigskip Possible figures:
\be
\item  M87 three-rail image of Hada: Figure 2 of
  http://www.mdpi.com/2075-4434/5/1/2

\item Laing and Bridle jet/counterjet velocity field analysis: top 4
  panels of figure 7 of

 https://academic.oup.com/mnras/article/424/2/1149/998560



\ee
}

\section{Bending}
\begin{figure}[h]
  \begin{center}$
\begin{array}{lcr}
    \includegraphics[width = 1.5in,angle = 0]{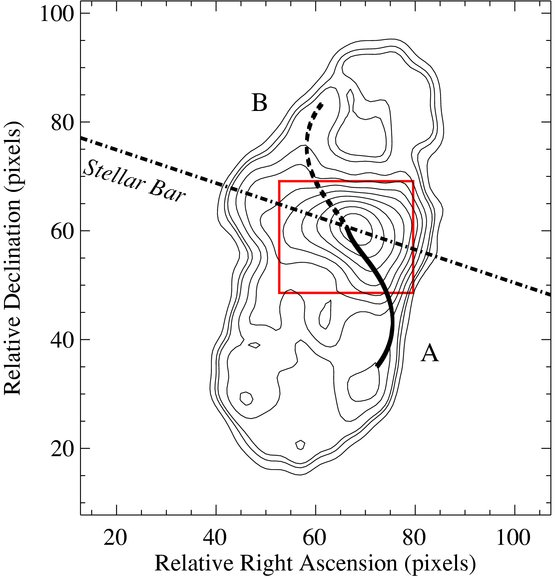} &
    \includegraphics[width = 1.5in,angle = 0]{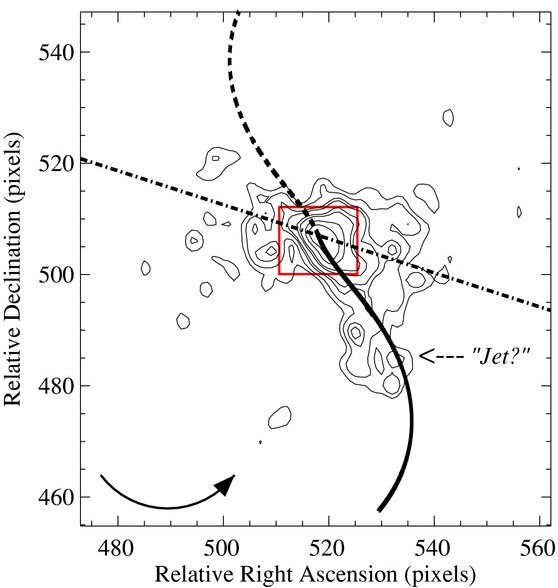} &    \includegraphics[width = 1.5in,angle = 0]{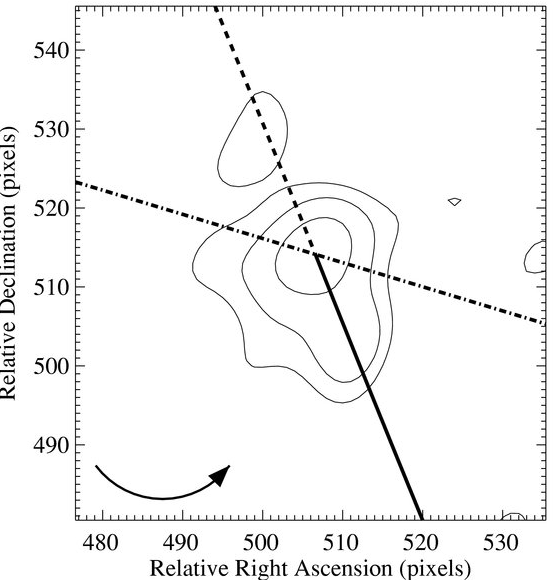} 

\end{array}$
\end{center}

\caption{\label{NGC6764}
  A precessing jet model fit to the radio emission from sub-kiloparsec
  to parsec scales in the Seyfert galaxy NGC~6764
  \citep{2010ApJ...723..580K}. The left panel shows a VLA 1.4 GHz
  image, the middle panel a VLA 5 GHz image, and the right panel a
  VLBA 1.6 GHz image. The red boxes indicate the region spanned by the
  next image panel in the sequence.  }

\end{figure}

Many jets have 'S'-shaped or more complex curvatures which have been
variously attributed to nozzle precession, hydrodynamic instabilities,
jet-cloud collisions, or ram pressure from the external medium.  By
tracing the jet at high resolution from pc to kpc scales, ngVLA
studies will help to distinguish between these scenarios for
individual objects, and determine for the first time the range and
form of bending in the AGN jet population (Fig.~\ref{NGC6764}). The
ngVLA will also probe a much larger variety of 'U'-shaped galaxies,
which have been used successfully to identify galaxy clusters and
directly study their intergalactic medium \citep{2017ApJ...844...78P}.
Opportunities will also exist for studying large scale hydrodynamic
instability modes in AGN jets, which manifest themselves as apparent
helical structures on $\sim 100$ pc scales
\citep{2005ApJ...620..646H}.  It will be possible to trace out to
large projected distances the extreme apparent bends in highly aligned
AGN jets (blazars), which on occasion directly cross our line of
sight, providing useful data for testing radiative transfer and jet
emission models \citep{2002ApJ...580..742H,1997AA...327..513A}.
 
\comment{
\bigskip Possible figures:
\be
\item 1345+125 helical jet:  Figure 6 of

  http://iopscience.iop.org/article/10.1086/345666/fulltext/

\item WAT image (NGC 1265) http://images.nrao.edu/19


\item 1510-089 Homan Fig 2:

 http://esoads.eso.org/abs/2002ApJ...580..742H
\ee
}
\section{Precession}
A recent discovery from long-term VLBA studies is that bright
superluminal features in pc-scale AGN jets will frequently emerge in a
particular direction that changes slowly on decadal timescales
\citep{2013AJ....146..120L}. The net result is that at any given time,
only the 'energized' portion of the jet containing recently ejected
features is visible with current interferometers. Over time,
subsequent ejections 'fill in' the jet, revealing its true conical
shape in stacked images \citep{2017MNRAS.468.4992P}. Evidence of
helical-type jet morphology has been seen on kpc-scales with the JVLA
that is suggestive of precession, and a particularly intriguing class
of 'X-shaped' radio galaxies may represent cases where a dramatic
90$^\circ$ or more shift has occurred in the jet direction over time.
The ngVLA will be vital for connecting the dynamic morphology seen on
pc-scales with emission immediately downstream, and for identifying
cases where the jet direction is actively shifting.

\section{\label{propermotions}Proper Motions and Astrometry}
The VLBA is a remarkable tool for precision jet astrometry, being able
to measure the speeds and accelerations of thousands of individual jet
features on decade or more timescales. Large surveys are able to
directly witness jet kinematics through time-lapse imaging, and have
revealed that powerful AGN jets are still undergoing collimation and
acceleration on scales of tens to hundreds of parsecs downstream of
the central engine \citep{MOJAVE_XII,2017ApJ...846...98J}, and
occasionally in abrupt fashion \citep{2003ApJ...589L...9H}.  Recent
work with HST and JVLA data on two AGN has shown that this type of
astrometry can also be performed on kiloparsec scale jet features
\citep{2017Galax...5....8M}.  The unprecedented sensitivity and $> 50$
km baselines of the ngVLA, in combination with improved resolution
afforded by JWST, will extend these measurements to a much larger set
of more distant and more powerful AGN.  Such studies are crucial for
constraining numerical models that seek to understand how and where
jets decelerate \citep[e.g.,][]{2018NatAs...2..167G}.

\section{Knots and Hotspots} 
\begin{figure}[h]
  \begin{center}$
\begin{array}{c}
    \includegraphics[width = 3.5in,angle = 0]{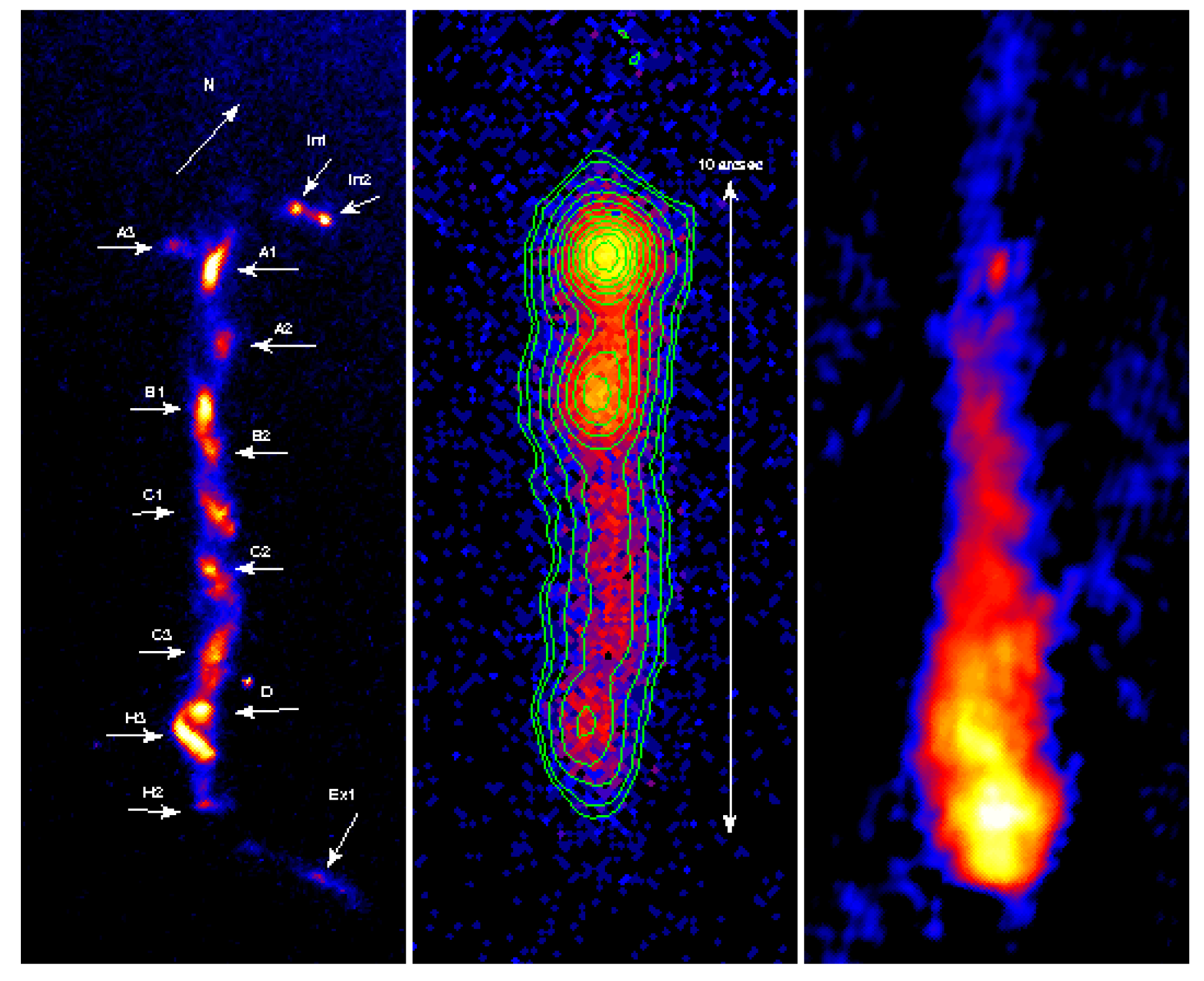} 

\end{array}$
\end{center}

\caption{\label{chandra3C273}
Images of the  30 kpc-long jet in the quasar 3C 273. From left to right, the images are optical (HST), X-ray (Chandra), and radio (MERLIN).
(Credit: Optical: NASA/STScI X-ray: NASA/CXC Radio: MERLIN) 
http://chandra.harvard.edu/photo/2000/0131/0131\_multi.tif
}
\end{figure}

Considerable uncertainties still exist
in our understanding of the emission processes in bright jet knots in
kpc-scale jets (Fig.~\ref{chandra3C273}). Recent multi-wavelength
studies with the JVLA, ALMA, HST, and Chandra have shown that simple
inverse-Compton scattering models with single-electron populations
cannot adequately account for the spectral energy distributions of jet
knots \citep[e.g.,][]{2017ApJ...835L..35M,2006ApJ...648..900J}.  In
some cases, such as M87, jet knots have been postulated to be the
sites of variable gamma-ray emission, but detailed supporting data are
lacking. Multi-epoch, multi-frequency polarimetric imaging with the
longest ngVLA baselines will make it possible to perform spectral
ageing, Faraday depth and magnetic field analyses of jet knots with
unprecedented resolution and sensitivity.  The terminal hotspots of
AGN jets, where the flow undergoes abrupt deceleration while
interacting the intergalactic medium, represent important targets for
ngVLA studies of feedback mechanisms, and for resolving why some AGN
display hybrid jet morphologies (Fig.~\ref{Kharbfig1}). In hybrid AGN,
one jet appears to be of high power, with a terminal hotspot, while
the other has properties more consistent with a lower power flow that
is decelerated gradually along its length by entrainment with the
external medium \citep{2000AA...363..507G}.  Hotspots and radio lobes
are also useful for probing the past activity history of the jets,
especially at longer radio wavelengths which sample the older electron
population.

\section{Radio-Loud and Radio-Quiet Quasars}
 
Fifty years after the discovery of quasars, the distinction between
radio-loud quasars (RLQ) and radio quiet quasars (RQQ) is still being
debated. Some authors have reported a clear bi-modal separation in
radio/optical luminosity ratio $R$ while others have claimed that the
distribution is continuous with no evidence for two populations (see
\citealt{2016ApJ...831..168K} for a review). Using the JVLA,
\cite{2011ApJ...739L..29K} and \cite{2016ApJ...831..168K} have
detected radio emission from essentially all 176 quasars in a
volume-limited sample from the SDSS contained within a narrow range of
red shift. Only about 10-15\% of the SDSS quasars fall in the
radio-loud classification, irrespective of whether the RLQs are
defined by $R$ or by luminosity. \cite{2013ApJ...768...37C} have shown
that the radio emission from the RQQ is due to star formation in the
host galaxy and is not directly related to the SMBH which gives rise
to the AGN. All quasars are thought to harbor a SMBH needed to support
the huge optical-IR luminosity which defines the category of quasars.
Why then is only a small fraction, 10-15\% of quasars RL?  Many
explanations have been offered, including intermittent activity,
absorption of the radio emission by intervening medium, host galaxy
properties, magnetic fields, and black hole spin or accretion rate.  A
particularly elegant interpretation was suggested by
\cite{1979Natur.277..182S} that RLQ are just the subset of quasars
whose relativistic beams are oriented nearly along the line of sight.
But at the time, it appeared that too large a fraction of optically
selected quasars appeared radio loud and that the observed detection
rate as a function of decreasing flux density was not consistent with
simple beaming models. Using more modern radio observations of
optically selected quasars and now knowing that relativistic jets
often have finite width and significant bends, as well as recognizing
that optically selected quasars are not necessarily randomly oriented
in the sky, the observed fraction (10-15\%) of RLQ and the observed
number flux density relations now appear consistent with relativistic
beaming models.  However, the extended radio emission seen in more
than half of RLQs represents a clear challenge to the relativistic
beaming interpretation of the RL/RQ dichotomy. Kinematic observations
of kpc (arcsec) scale are difficult as the expected motions are
typically only a few tenths of a milliarcsec per year. Currently there is
evidence of modest superluminal motion in 7 mm observations of M87
\citep{2018arXiv180206166W}, and in the kpc-scale jets of a few AGN (
\S~\ref{propermotions}). High-resolution-high sensitivity proper motion
observations with the ngVLA will give further insight to the importance of
relativistic beaming in kpc scale radio jets and its implications for
the RL/RQ dichotomy.

\section{Low-luminosity/Radio-quiet AGN}

More than 80\% of AGN, comprising Seyfert and Low Ionization Nuclear
Emission Line Region (LINER) galaxies, do not possess collimated
plasma outflows on kpc-scales. Weak AGN outflow emission is often
indistinguishable from outflows due to galactic stellar activity or
nuclear starbursts \citep[e.g.,][]{2006AJ....132..546G}.
Disentangling AGN and stellar contributions to the radio emission are
essential to understanding the nature of the AGN in low luminosity AGN
(LLAGN), and in discerning why the vast majority of AGN do not produce
large-scale jets. Multi-scale observations from parsec to kpc-scales
can trace the connection between an AGN jet and kpc-scale lobes
observed in many LLAGN (Fig.~\ref{NGC6764}). Polarization-sensitive
imaging can further distinguish between the stellar and AGN
outflows, due to presumably more organized magnetic field structures
and higher degrees of polarization in AGN outflows compared to stellar
outflows. The ngVLA will detect fainter radio
emission in LLAGN and in doing so will trace AGN outflows as they are
launched from the black hole-accretion-disk systems and propagate
through the ISM of their host galaxies.

\bigskip


\bibliography{lister}  


\end{document}